\documentclass{PoS}
\usepackage{latexsym}
\usepackage{amssymb}
\usepackage{amsmath}

\title{Preliminary study of two-dimensional SU(N) Yang-Mills theory with adjoint matter by Hybrid Monte Carlo approach}

\ShortTitle{D=2 SU(N) Yang-Mills theory with adjoint matter}

\author{\speaker{Piotr Korcyl}\\
        Jagiellonian University, Institute of Physics, ul. Reymonta 4, 30-059 Krak\'ow, Poland\\
        E-mail: \email{piotr.korcyl@uj.edu.pl}}

\author{Mateusz Kore\'n\\
        Jagiellonian University, Institute of Physics, ul. Reymonta 4, 30-059 Krak\'ow, Poland\\
        E-mail: \email{mateusz.koren@uj.edu.pl}}

\abstract{Two-dimensional non-abelian quantum field models provide a useful laboratory
for analytic and numerical investigations of quantum theories with gauge symmetry. They
can exhibit various features, such as charge confinement, which are known from D=4 theories like QCD.
Several analytic predictions concerning the spectra of two-dimensional systems with adjoint matter were postulated
and numerical results were obtained using Discrete Light Cone Quantization techniques, however none of them has been checked
via Monte Carlo simulations.
In this Letter we present two such models which are particularly interesting from the physical point of view
and discuss first numerical results.
}

\FullConference{ The XXIX International Symposium on Lattice Field Theory - Lattice 2011\\
July 10-16, 2011\\
Squaw Valley, Lake Tahoe, California}

\begin{document}


Grasping the full, nonperturbative dynamics of strong interactions is a difficult task. Although current cutting-edge lattice
simulations of QCD provide the hadron spectrum obtained from first principles \cite{bmw}, they do not unravel the hidden
complexity of the theory. Hence, one would like to simplify QCD in order to disentangle some of the mechanisms at play
and investigate them separately. From this point of view, two ways of approximating QCD appear as especially
appealing. The first one is dimensional reduction, where one investigates lower dimensional theories instead of $D=4$ ones.
Although the former are much simpler, they may still possess some of intrinsic features of the original theory.
The second way is the approximation which uses the large-N limit \cite{thooft1}. In this case, one replaces the $SU(3)$ color symmetry group by
$SU(N)$ and studies the theory in the limit of $N \rightarrow \infty$. An interplay of both these approximations enabled to construct several
models having interesting properties. In the following we will describe two of them, which appear particularly attractive, and
present some numerical results obtained for systems which are related, but technically simpler to simulate.

\section{Models of QCD}

\subsection{Fundamental matter}

In 1962 Julian Schwinger solved analytically a two-dimensional version of quantum electrodynamics \cite{schwinger}.
He showed that massless electrons and positrons are confined into neutral pairs - massive mesons.
In this model a free charge of any value is screened by pairs coming from the vacuum \cite{susskind}.
Hence, the spectrum is composed exclusively of identical, massive mesons as is clear from the bosonised version of the model.
Subsequently, generalizations for systems with $N_f$ flavors of electrons were found \cite{hetrick}, in case of which
one gets, with a nomenclature borrowed from QCD, a number of massless pions and a single massive eta meson.
No exact solutions exist for models with massive electrons, however several approximate results were given \cite{hetrick,bergknoff}.
Non-abelian models were also considered \cite{gross}. For the $SU(3)$ gauge symmetry group, such system corresponds simply to dimensionally reduced QCD.

\begin{figure}
\begin{center}
\includegraphics[width=0.15\textwidth, angle=270]{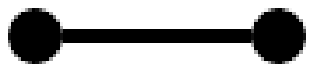}
\hspace{2cm}
\includegraphics[width=0.15\textwidth, angle=270]{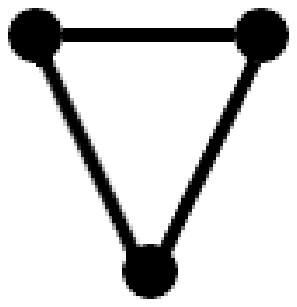}
\end{center}
\caption{Schematic plot showing the difference between bound states composed of fundamental (left) and adjoint (right) matter (dots) and glue fluxes (segments). Fundamental matter
fields have a single color index; in the large $N$ limit they can be bound in two body states only. Adjoint matter fields have two color indices; hence, many body bound
states are possible.}
\label{fig. schematic bound states}
\end{figure}

In 1974 't Hooft proposed another theoretical tool to simplify QCD. He considered a two dimensional
version of the theory in the limit of large number of colors \cite{thooft2} and found that the complexity
of the dynamics is drastically reduced when $N$ is taken large.
The resulting spectrum had several phenomenologically attractive features such as narrow,
non-interacting mesons (see figure \ref{fig. schematic bound states}).
However, this approximation had also some drawbacks: quark loops were suppressed, hence rendering
many body bound states invisible and problems with reconstruction of barionic spectrum appeared.

From the numerical perspective a lot of work was devoted to the Schwinger model. On one hand, due to its
technical simplicity it was used as a testbed for many numerical algorithms \cite{banks_susskind_kogut,neuberger}. On the other hand,
due to the relative complexity of its spectrum and of phenomena present in it, the Schwinger model gave a lot
of understanding into the dynamics of confining gauge theories. Many complementary numerical results were obtained through different
lattice calculations as well as with the discrete light cone quantization method (for the latter see for example Ref.\cite{brodsky}). 

\subsection{Adjoint matter}

It seems very likely that the interplay between a lower-dimensional theory and the large-N limit can lead to a model
which can capture some essential features of QCD. A way to correct the drawbacks of the 't Hooft model is to change
the representation of the matter fields. One can use quarks in the antisymmetric representation \cite{armoni_veneziano}, which
for $N=3$ reduces to the fundamental one. Hence, one gets a theory which corresponds to QCD for $N=3$ but possess
a different large-N limit, i.e. quark loops are not negligible and the spectrum may contain many body bound states.
Another possibility is to use adjoint matter fields. In this case, quark loops are of the same order as the gluonic ones in
the expansion parameter $1/N$ for $N$ large.

The two most common ways to introduce to the model matter fields in the adjoint representation are either by dimensional
reduction or by supersymmetry. In the former case one gets scalar fields as the components of the reduced gauge potential, whereas in the
latter case one obtains fermions as the superpartners of the bosonic gauge degrees of freedom. Both possibilities 
were investigated in Refs.\cite{dalley_klebanov}.

The simplest model, which also attracted some attention, is composed of a single flavor of adjoint Majorana fermions \cite{dalley}.
Its spectrum contains many body bound states and an exponentially growing density of states - characteristic features of QCD spectrum.
As far as the numerical results are concerned, only findings obtained
by the discrete light front quantization technique are available \cite{dalley_numerics} - no lattice simulations
of this two dimensional model were performed up to now.


Therefore, we set up a project of a systematic study by lattice methods of QCD$_2$ models with adjoint Majorana fermions for different $SU(N)$
gauge groups. Our aim is to crosscheck the discrete light cone quantization method results and verify theoretical predictions.
We have chosen two particular models which were not studied on the lattice extensively and for which a connection to QCD can be argued.
We briefly describe these two systems in the next section.

\section{Interesting models}

\subsection{Kutasov model}

In Ref.\cite{kutasov} Kutasov considered a system consisting of a $SU(N)$ gauge field and $N^2-1$ real Majorana
fermions transforming in the adjoint representation of the gauge group, described by means of a traceless, hermitian $N \times N$ matrix.
The spectrum is expected to consist of many body bound states (see figure \ref{fig. schematic bound states})
 with a density of states of a given mass raising exponentially with mass \cite{dalley_klebanov}.
The lagrangian is given by,
\begin{equation}
\mathcal{L} = \textrm{Tr}\Big( \frac{1}{g^2} F_{\mu \nu}^2 + \sum_{k=1}^{N_f} \big( \bar{\psi}^k \gamma^{\mu} D_{\mu} \psi^k + m_k \bar{\psi}^k \psi^k \big) \Big).
\label{eq. kutasov model}
\end{equation}
with $g$ and $m$ being two dimensionfull parameters of the dimension of mass.

In two space-time dimensions there is no dynamical gauge degrees of freedom, hence
the model contains only fermionic degrees of freedom.
An analytic calculation in the light cone coordinate frame can reveal this.
In the $A_-^{i j}=0$ gauge the Lagrangian reads
\begin{equation}
\mathcal{L} = \textrm{Tr}\Big( \frac{1}{g^2} \big( \partial_- A_+ \big)^2 + i \psi \partial_+ \psi + i \bar{\psi} \partial_- \bar{\psi}-2i m \bar{\psi} \psi + A_+ J^+ \Big),
\end{equation}
where $\psi_{i j}$ and $\bar{\psi}_{i j}$ denote respectively the right and left moving fermions. After integrating the non dynamical
degrees of freedom one gets
\begin{equation}
\mathcal{L} = \textrm{Tr}\Big(i \psi \partial_+ \psi + g^2 J^+ \frac{1}{\partial_-^2} J^+ - i m^2 \psi \frac{1}{\partial_-} \psi \Big),
\label{eq. kutasov light front}
\end{equation}
which is exclusively composed of right-moving fermions.

On first sight the system eq.\eqref{eq. kutasov light front} cannot be supersymmetric, since the numbers of bosonic and fermionic degrees of freedom.
Nevertheless, it turns out that for a particular value of the quotient $g/m$ the model possess a \emph{dynamical} supersymmetry \cite{kutasov}.
Namely, there exist two supersymmetry generators given by
\begin{align}
Q^+ &= \frac{1}{3\sqrt{N}} \int_{-\infty}^{\infty} dp \ dq \psi_{i j}(-p) \psi_{j k}(q) \psi_{k i}(p-q), \nonumber \\
Q^- &= g \int_{-\infty}^{\infty}  \frac{dp \ dq}{p} \psi_{i j}(-p) \psi_{j k}(q) \psi_{k i}(p-q),
\end{align}
satisfying for $m^2 = g^2 N$
\begin{align}
\big[ Q^+, P^- \big] = 0, &\qquad \big[ Q^-, P^- \big] = 0, \nonumber \\
\big( Q^+ \big)^2 = P^+, &\qquad \big( Q^- \big)^2 = - P^-,
\end{align}
$P^-$ and $P^+$ being the light-cone Hamiltonian and total momentum operators respectively. Moreover
\begin{equation}
\big\{ Q^+, Q^- \big\} = 0.
\end{equation}


Finally, one should note a result published in Ref.\cite{kutasov2} pointing to a partial equivalence between the model eq.\eqref{eq. kutasov model}
and a model with fundamental matter. Namely, the Authors proved that the massive sector of a system with $SU(N)$ color symmetry and a single adjoint
fermion is equivalent to the massive sector of a system with $SU(N)$ color symmetry and $N$ flavors of fundamental fermions.

\subsection{$\mathcal{N}=(1,1)$ model}

The second model of interest is a two-dimensional theory obtained by dimensional reduction of
the $D=4$, $\mathcal{N}=1$ supersymmetric Yang-Mills theory down to $D=2$. It was
considered as a toy model of QCD by Dorigoni, Wosiek and Veneziano \cite{wosiek}, where
it was analyzed numerically by the light-cone quantization method. This system was also studied
on the lattice in Ref.\cite{hanada} in the context of supersymmetry on the lattice.

The $D=4$ theory reduced to $D=2$ yields a lagrangian of the following form,
\begin{multline}
\mathcal{L} = \textrm{Tr} \Big(-\frac{1}{4} F_{\mu \nu}F^{\mu \nu} + i \lambda^{\dagger} \gamma^{\mu} D_{\mu} \lambda
+ i \chi^{\dagger} \gamma^{\mu} D_{\mu} \chi + \\ + \frac{1}{2} D_{\mu} \phi_2 D^{\mu} \phi_2 + \frac{1}{2} D_{\mu} \phi_3 D^{\mu} \phi_3 + \textrm{interation terms}\Big).
\end{multline}
Hence, it may be viewed as a generalization of the Kutasov model by inclusion of a second fermion and two scalar fields.

Using the numerical light-cone approach and the Coulomb approximation the model was shown to be a
non-trivial generalization of the 't Hooft's model with an arbitrary number of partons. The spectrum turned out to be composed
of color-less bound states of many partons whose mass grows linearly with distance
between the partons, namely $M \sim \sigma \sum |\Delta x|$ \cite{wosiek}. Hence, a string picture of flux tubes connecting
each parton emerged.


\section{First results and plots}

\begin{figure}
\begin{center}
\includegraphics[width=0.33\textwidth, angle=270]{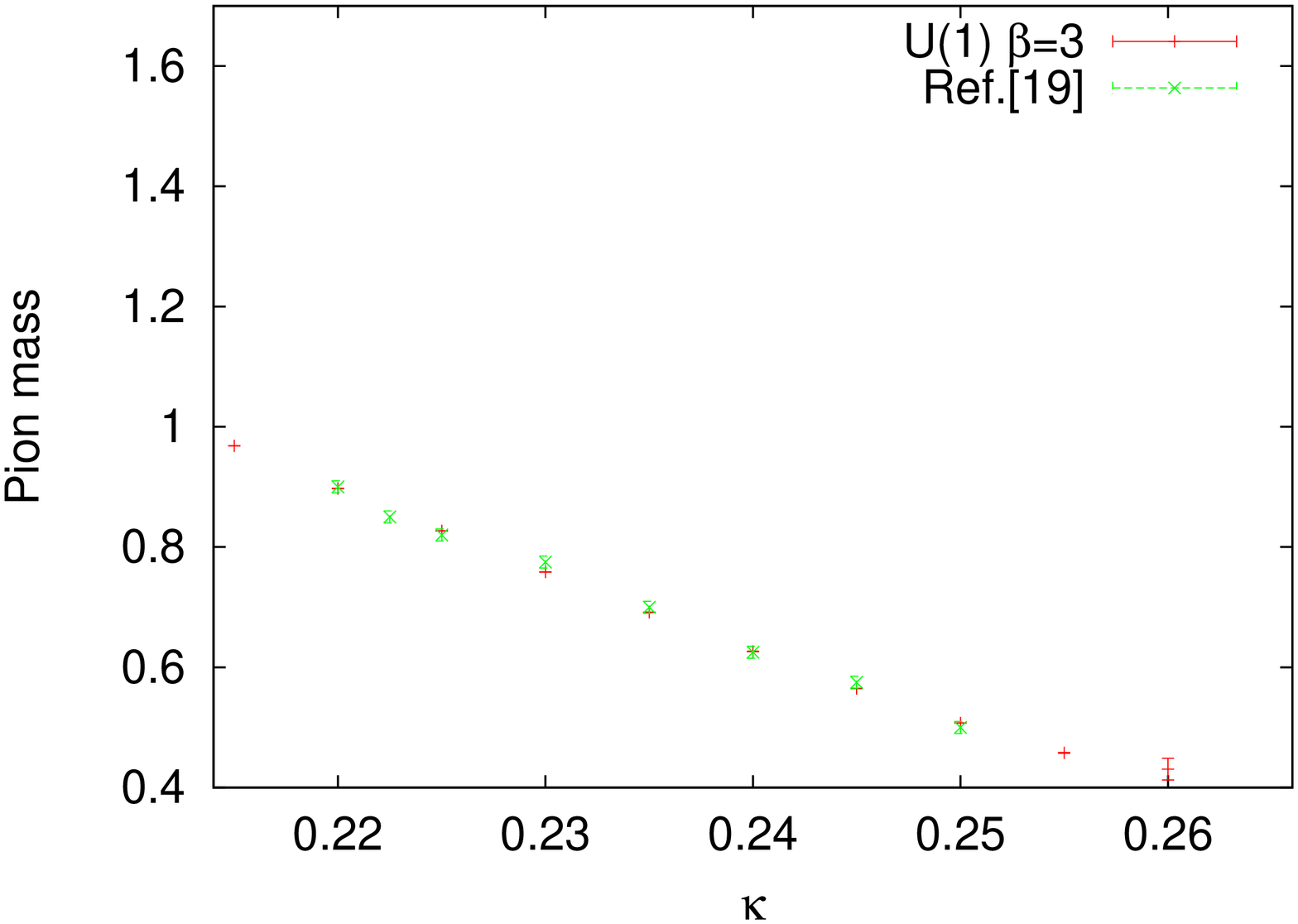}
\includegraphics[width=0.33\textwidth, angle=270]{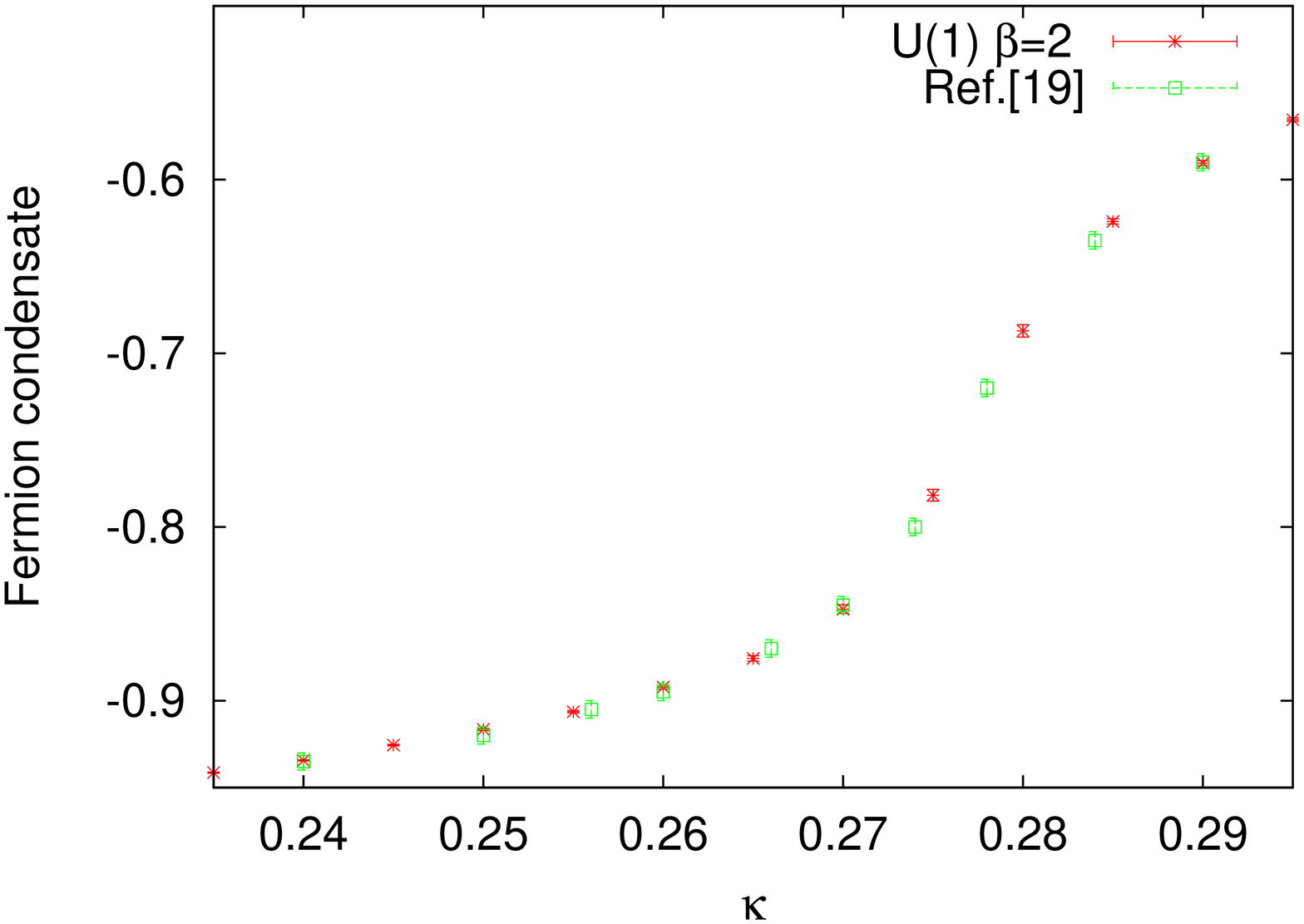}
\end{center}
\caption{Pion mass and fermion condensate for the Schwinger model, comparison with Ref.\cite{elser} (parameters: 8x20 lattice, $10^4$ measurements).}
\label{fig. elser}
\end{figure}

\begin{figure}
\begin{center}
\includegraphics[width=0.33\textwidth, angle=270]{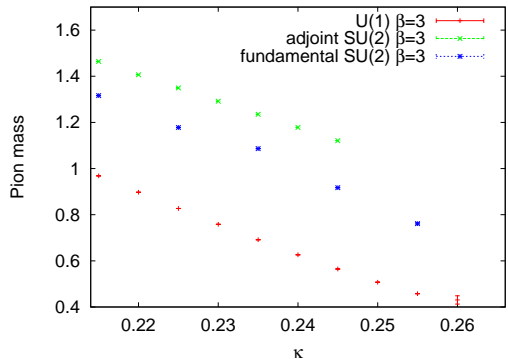}
\end{center}
\caption{Pion masses for $U(1)$ and $SU(2)$ fundamental and adjoint models (parameters: 8x20 lattice, $10^4$ measurements).}
\label{fig. adjoint}
\end{figure}

We have started our project by setting up simulations with two Dirac adjoint fermions. We have implemented dynamical
fundamental and adjoint
fermions using the Hybrid Monte Carlo algorithm. A first set of tests consisted in reproducing the known results
for the Schwinger model. On figure \ref{fig. elser} we
compare our results with the numbers published in Ref.\cite{elser}. A very good agreement is seen. Then, we have performed
simulations for the $SU(2)$ gauge group. Figure \ref{fig. adjoint} show the comparison of the
pion mass for the $U(1)$, $SU(2)$ fundamental and $SU(2)$ adjoint models.
The equivalence of Ref.\cite{kutasov2} and conclusions of Ref.\cite{hetrick}
suggests that the massive states
of the $SU(2)$ adjoint model should be $\sqrt{2}$ times heavier than the massive states of the $SU(2)$ fundamental model.
Data shown on figure \ref{fig. adjoint} indeed hint that the adjoint pions are heavier than the fundamental ones. However
for the exact determination of the coefficient studies in the $\eta$ sector are needed and extrapolations
to the continuum and to the chiral limit must be performed.

%
%
%




\section{Conclusions}

Summarizing, we argued that two-dimensional Yang-Mills theories
enable to grasp particular features of quantum chromodynamics. By appropriately choosing the representation
of fermionic matter fields it is possible to obtain models with an interesting large-$N$ limit.
Through systematic studies one may find a system which resembles QCD particularly well,
but thanks to its simplified structure may be analytically tractable and lead to a better understanding
of the mechanisms at play. Lattice approach
is particularly well suited for such investigations.
We recalled two two-dimensional models, a supersymmetric Yang-Mills theory
obtained by dimensional reduction, and a non-supersymmetric Yang-Mills theory with adjoint matter fields
which turns out to be supersymmetric for certain value of parameters, as particularly interesting to study numerically.
Finally, we discussed first numerical results. Further studies are on the way.


\end{document}